\documentclass[aps,prl, reprint, superscriptaddress,preprintnumbers,amssymb]{revtex4-2}
\usepackage{graphicx}
\usepackage{dcolumn}
\usepackage{bm}

\begin{document}

\preprint{ver3}

\title{Anomalous upper critical field in the quasicrystal superconductor Ta$_{1.6}$Te}



\author{Taichi Terashima}
\email{TERASHIMA.Taichi@nims.go.jp}
\affiliation{Research Center for Materials Nanoarchitectonics (MANA), National Institute for Materials Science, Tsukuba 305-0003, Japan}
\author{Yuki Tokumoto}
\email{tokumoto@iis.u-tokyo.ac.jp}
\author{Kotaro Hamano}
\affiliation{Institute of Industrial Science, The University of Tokyo, Tokyo 153-8505, Japan}
\author{Takako Konoike}
\affiliation{Research Center for Materials Nanoarchitectonics (MANA), National Institute for Materials Science, Tsukuba 305-0003, Japan}
\author{Naoki Kikugawa}
\affiliation{Center for Basic Research on Materials, National Institute for Materials Science, Tsukuba 305-0003, Japan}
\author{Keiichi Edagawa}
\affiliation{Institute of Industrial Science, The University of Tokyo, Tokyo 153-8505, Japan}


\date{\today}

\begin{abstract}
Superconductivity in quasicrystals poses a new challenge in condensed matter physics.
We measured the resistance and ac magnetic susceptibility of a Ta$_{1.6}$Te dodecagonal quasicrystal, which is superconducting below $T_c \sim$ 1 K.
We show that the upper critical field increases linearly with a large slope of $-$4.4 T/K with decreasing temperature down to 0.04 K, with no tendency to level off.
The extrapolated zero-temperature critical field exceeds the Pauli limit by a factor of 2.3.
We also observed flux-flow resistance with thermally activated behavior and an irreversibility field that is distinct from the upper critical field.
We discuss these peculiarities in terms of the nonuniform superconducting gap and spin--orbit interaction in quasicrystal structures.
\end{abstract}


\maketitle



%

Quasicrystals (QCs), first reported by Shechtman \textit{et al}. in 1984 \cite{Shechtman84PRL}, lack a periodic structure (translational symmetry), but their diffraction patterns exhibit sharp Bragg spots, which indicate rotational symmetries such as five-, eight-, ten-, or twelve-fold symmetry, which are forbidden in periodic crystals.
Since their discovery, many quasicrystals have been synthesized \cite{Tsai13CSR} and some natural quasicrystals have also been found \cite{Bindi09Science}.

Because many basic concepts in solid-state physics rely on lattice periodicity and its associated Brillouin zone, the possibility of long-range electronic order in quasicrystals is an intriguing prospect. 
Two studies have reported affirmative results:
Kamiya \textit{et al}. discovered superconductivity in Al--Zn--Mg icosahedral quasicrystals (i-QC) \cite{Kamiya18NatCommun}, while Tamura \textit{et al}. reported long-range magnetic order in Au--Ga--Gd and Au--Ga--Tb i-QCs \cite{Tamura21JACS}.

The report of Kamiya \textit{et al}. and earlier works \cite{Graebner87PRL, Wong87PRB, Wagner88PRB, Azhazha02PhysLettA} have inspired theoretical investigations of quasicrystal superconductivity. 
Sakai \textit{et al}., who studied an attractive Hubbard model on a Penrose lattice \cite{Sakai17PRB}, found a superconducting state with a spatially inhomogeneous superconducting gap.
They showed that Copper pairs are spatially extended in the weak-coupling regime.
Similar superconducting states with a spatially nonuniform gap are reported elsewhere \cite{Araujo19PRB, Liu22SCPM}.
A nonuniform gap in quasicrystal superconductors can lead to peculiar superconducting properties. 
For instance, it suppresses the Bogoliubov quasiparticle peak in the density of states, possibly causing unconventional current--voltage ($I-V$) characteristics \cite{Sakai19PRR}, and can impose intrinsic vortex-pinning sites \cite{Nagai22PRB}.
In addition, Sakai \textit{et al}. argued that an exotic superconducting state with a spatially sign-changing order parameter, which is reminiscent of the Fulde--Ferrell--Larkin--Ovchinnikov (FFLO) state, may emerge at low temperatures and high fields in the temperature--field phase diagram of quasicrystal superconductors \cite{Sakai19PRR}.

The low superconducting transition temperature $T_c$ of the Al--Zn--Mg i-QC ($\sim$0.05 K) impedes detailed investigations of the superconducting properties of this quasicrystal.
Recently, Tokumoto \textit{et al}. reported a Ta$_{1.6}$Te dodecagonal quasicrystal (dd-QC) superconductor with a much higher $T_c$ (0.98 K) \cite{Tokumoto23condmat}, enabling studies of quasicrystal superconductivity with various probes.

The Ta$_{1.6}$Te dd-QC is a layered material in which $\sim$1-nm-thick Te-terminated layers are separated by van der Waals (vdW) gaps \cite{Conrad98AngewChemIntEd, Conrad02ChemEurJ, Cain20PNAS}.
Tokumoto \textit{et al}. demonstrated superconducting properties, namely, zero resistivity, the Meissner effect, and a specific-heat jump, in the Ta$_{1.6}$Te dd-QC \cite{Tokumoto23condmat}.
Combining the McMillan formula with specific-heat data, they concluded a weak-coupling superconductivity with an electron--phonon coupling constant $\lambda_{ep}$ of 0.52. 
They also measured the upper critical field down to $T/T_c \sim$ 0.4.
At the lowest temperature ($T$ = 0.43 K), the $B_{c2}$ reached 2.3 T, exceeding the Pauli limit (paramagnetic critical field) of magnetic-field strength  $B_{po}$ = 1.8 T, where superconductivity is expected to be destroyed by Zeeman-energy gain of the electron spins.
Within the framework of weak-coupling Bardeen--Cooper--Schrieffer (BCS) theory, $B_{po}$ (in Tesla) is given by 1.84 $T_c$ (in Kelvin)  \cite{Decroux82Book}.
The temperature dependence of $B_{c2}$ was almost linear in the measured range, but it could also be described by the standard Werthamer--Helfand--Hohenberg (WHH) theory \cite{Werthamer66PRB} because of the limited temperature range.
The WHH computation of $B_{c2}(T)$ applies the weak-coupling BCS theory to a spherical Fermi surface.

In this study, we measure the resistance $R$ and ac magnetic susceptibility $\textrm{ac-}\chi$ ($\chi^{\prime} - i\chi^{\prime\prime}$) of the Ta$_{1.6}$Te dd-QC down to 0.04 K ($T/T_c$ = 0.04) and illuminate peculiarities in the superconductivity of this quasicrystal.
Especially, we show that the upper critical field $B_{c2}(T)$ increases linearly with decreasing temperature down to 0.04 K without leveling off, and that the estimated zero-temperature critical field $B_{c2}(0)$ far exceeds the Pauli limit.


\begin{figure}
\includegraphics[width=8.6cm]{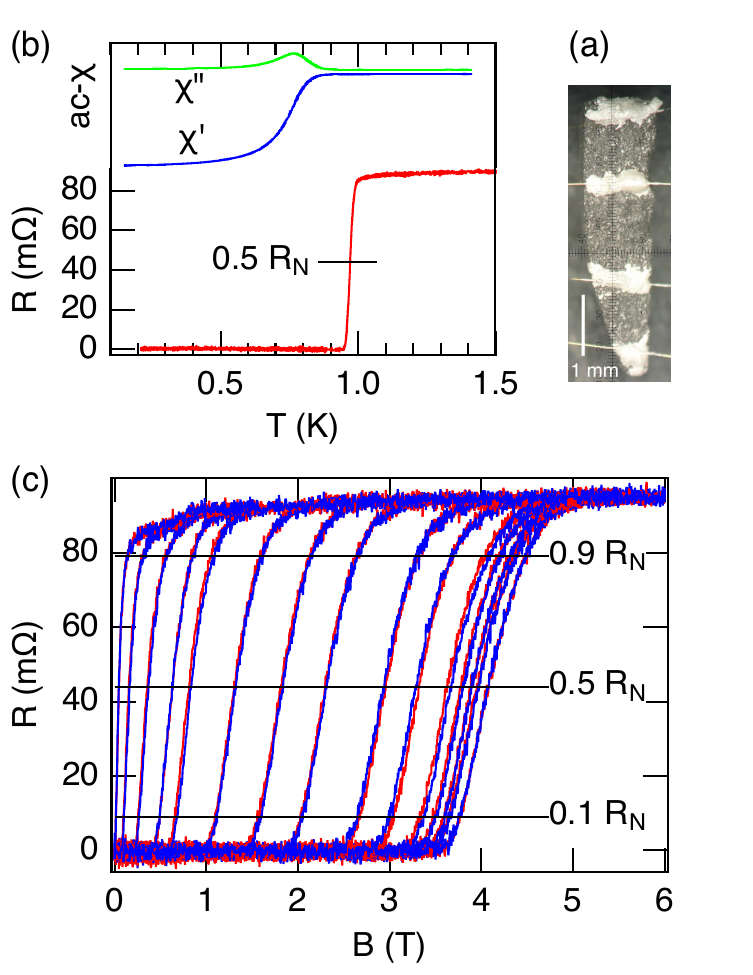}
\caption{\label{transition}Superconducting transition in the Ta$_{1.6}$Te dd-QC.
(a) Photograph of the sample.
(b) Temperature dependences of resistance $R$ and ac magnetic susceptibility $\textrm{ac-}\chi$ ($\chi^{\prime} - i\chi^{\prime\prime}$).
(c) Magnetic-field dependence of resistance.  
At each set temperature (0.039, 0.054, 0.067, 0.092, 0.123, 0.219, 0.294, 0.433, 0.532, 0.642, 0.756, 0.802, 0.866, 0.920, and 0.956 K from right to left), the field was swept up (red) and down (blue).
}
\end{figure}

\begin{figure}
\includegraphics[width=8.6cm]{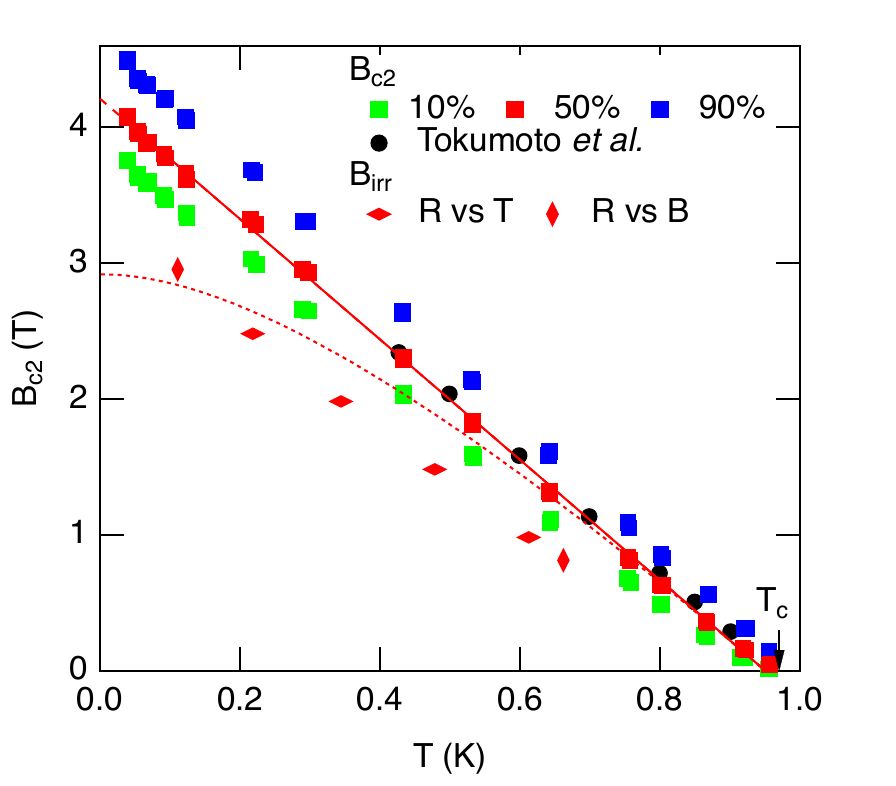}
\caption{\label{Bc2}Superconducting phase diagram of the Ta$_{1.6}$Te dd-QC.
The upper critical field $B_{c2}$ data determined under the 10\%, 50\%, and 90\%  criteria (squares) are compared with the
$B_{c2}$ data of Tokumoto \textit{et al}. \cite{Tokumoto23condmat} (circles).
The straight line is fitted to the 50\% $B_{c2}$ data and the dotted line describes the orbital critical field without the Pauli limit based on WHH theory. 
The irreversibility fields $B_{irr}$ determined from Figs. 3(b ) and 4(a) are also shown (upward and sideways diamonds, respectively).
}
\end{figure}

The study was performed on a polygrain sample [Fig. 1(a)] prepared via reaction sintering of TaTe$_2$ and Ta (for materials and methods, see the Supplemental Material \cite{SM_Ta1p6Te}).
The sample contains a small amount of superconducting impurity, which causes a resistivity drop of $\sim$2\% at $T_c \sim 3.2$ K [Fig. S1(a)].
However, as the upper critical field is small [$\sim$0.5 T at $T$ = 1.88 K; see Fig. S1(b)], this impurity negligibly disturbs the present measurements.

We first examine the superconducting transition and upper critical field.
Figure 1(b) shows the temperature dependences of the resistance $R$ and ac magnetic susceptibility below 1.5 K.
The resistance at $T$ = 1.1 K is $R_N$ = 88 m$\Omega$, approximately corresponding to a resistivity $\rho$ of $\sim$4 m$\Omega$~cm.
Tokumoto \textit{et al}. \cite{Tokumoto23condmat} reported a $\rho$ of 1.7 m$\Omega$~cm at $T$ = 300 K, which increased by $\sim$10 \% during cooling of the sample to 4.2 K.
Considering the irregular sample shape [Fig. 1(a)], the two values fairly agree.
The resistance sharply drops below $\sim$1 K, signaling a superconducting transition.
The midpoint transition temperature is $T_c$ = 0.97 K, which favorably agrees with the value obtained by Tokumoto \textit{et al} (0.98 K).
The 10\%--90\% transition width is 0.03 K.
The real part $\chi^{\prime}$ of the ac magnetic susceptibility begins deviating from the normal-state value at $\sim$0.95 K, roughly corresponding to the temperature at which $R \rightarrow 0$ as usual.
The diamagnetic response confirms the superconductivity of the sample.

Figure 1(c) shows the $R$ versus $B$ curves measured at various temperatures.
At all set temperatures except 0.22 K, the temperature was stabilized within $\sim$2\% (at 0.22 K, the temperature stabilized within $\sim$5\%).  
The magnetic field was swept up and down at each temperature.
To avoid self-heating of the sample, the measurement current was reduced to 7 $\mu$A, corresponding to a current density of 1$\times$10$^{-3}$ Acm$^{-2}$.
For technical reasons, the field direction was offset by 20$^{\circ}$ from the parallel direction $B \parallel I$.
At the highest and second-highest set temperatures, 0.96 and 0.92 K, respectively, the resistance increased rapidly to $\sim$80 m$\Omega$, followed by a gradual rise until $B$ reached $\sim$0.8 T.
The gradual increase was caused by the impurity phase, which has a higher $B_{c2}$ than the main quasicrystal phase at these temperatures.
Lowering the temperature broadened the resistive transition: the 10\%--90\% transition width increased from $\Delta$$B$ = 0.12 T at 0.93 K to 0.72 T at 0.04 K.

We determined the upper critical field $B_{c2}$ from the $R(B)$ data under three criteria: 10\%, 50\%, and 90\% of the normal resistance $R_N$.
The resultant $B_{c2}$'s are plotted in Fig. 2.
The 50\% data well agree with those of Tokumoto \textit{et al}. \cite{Tokumoto23condmat}.
Under each criterion, the $B_{c2}$ versus $T$ plot was linear down to $T$ = 0.04 K ($t$ = $T/T_c$ = 0.04).
Straight-line fitting of the 50\% result yielded a slope d$B_{c2}$/d$T$ of $-4.43$(2) T/K and an extrapolated $B_{c2}(0)$ of 4.21(1) T, corresponding to a coherence length of $\xi$ = 88.4 \AA.
Note that this experimental $B_{c2}(0)$ exceeds the Pauli limit $B_{po}$ = 1.8 T by a factor of 2.3.
Applying WHH theory with the dirty limit and taking the above $B_{c2}$ slope as the initial slope of $B_{c2}$, $\mathrm{d}B_{c2}/\mathrm{d}T|_{T_c}$, we also calculated the theoretical $B_{c2}$ versus $T$ curve (dotted line in Fig. 2).
As the Pauli limit was neglected in this calculation, the calculated curve corresponds to the orbital critical field $B_{c2}^*$ and gives the largest possible upper critical field in WHH theory.
The experimental $B_{c2}$ surpasses the theoretical curve at low temperatures.

\begin{figure}
\includegraphics[width=8.6cm]{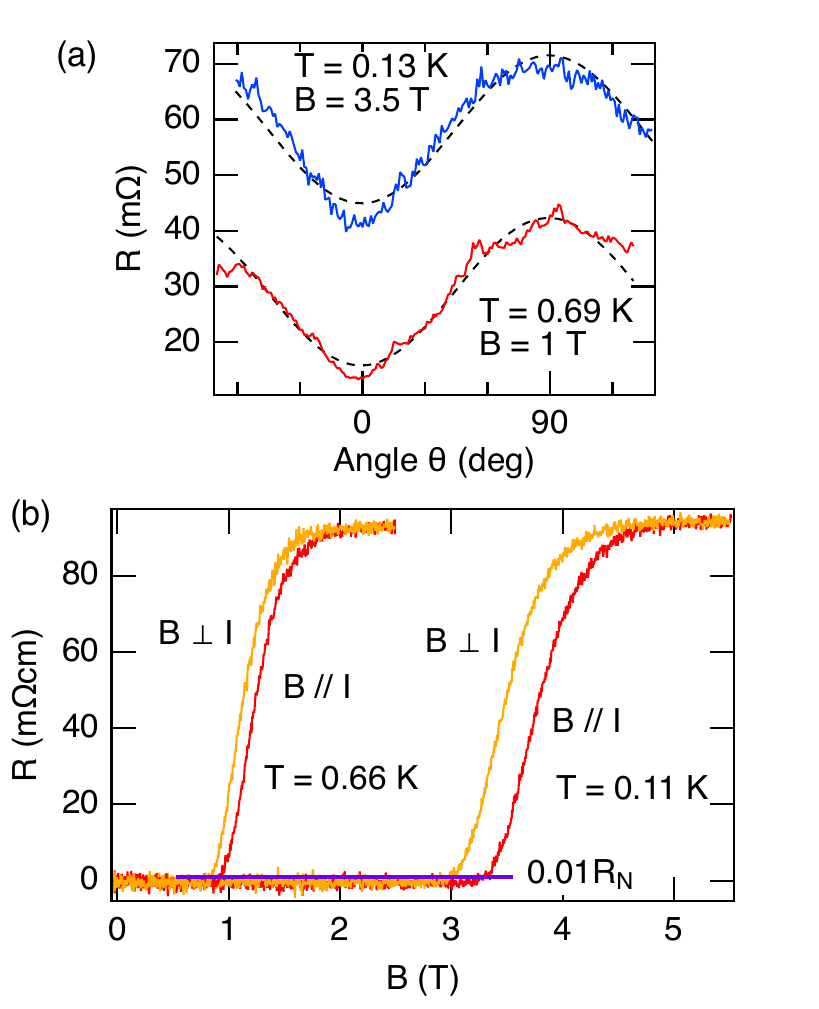}
\caption{\label{Flux-flow}Flux-flow resistance in the Ta$_{1.6}$Te dd-QC.
(a) Resistance as a function of the angle $\theta$ between the field and current.
Measurements were performed at the stated temperatures and fields in the resistive-transition region.
The broken lines are fitted to $\sin^2(\theta)$.
(b) Resistive-transition curves of $B \parallel I$ and $B \bot I$ at the indicated temperatures.
}
\end{figure}

\begin{figure}
\includegraphics[width=8.6cm]{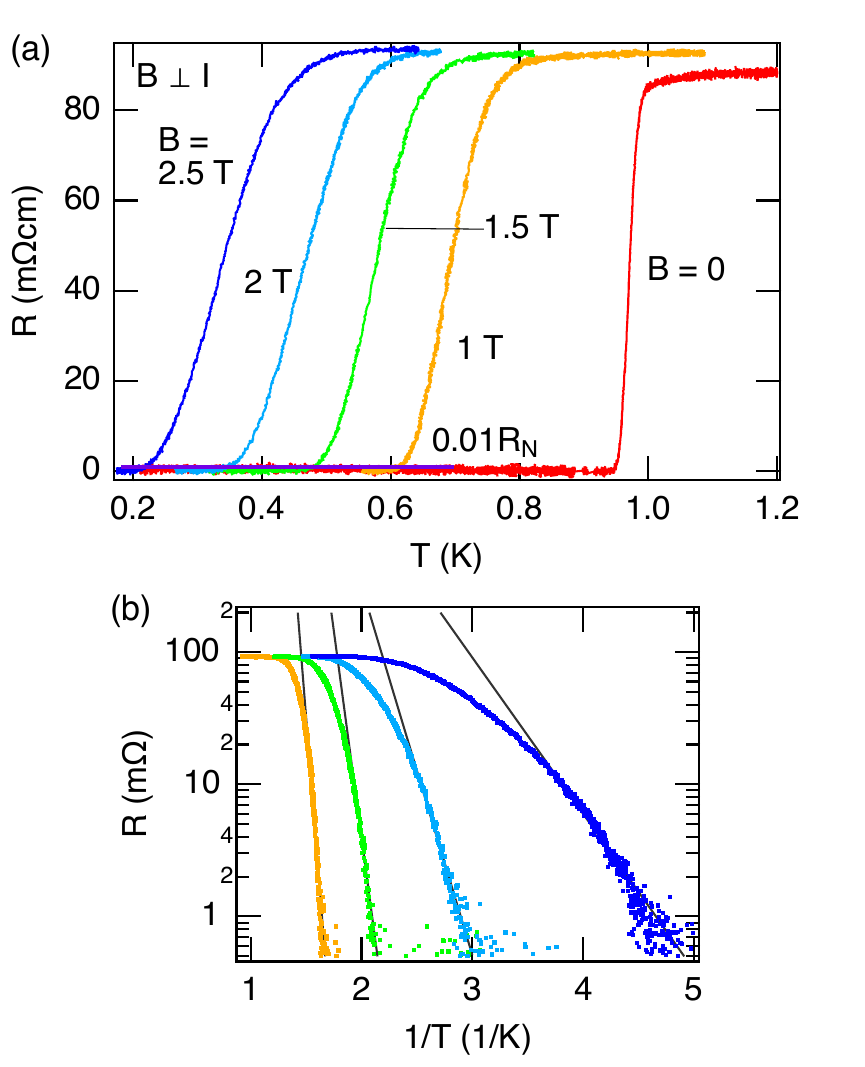}
\caption{\label{Bc2}Temperature dependence of the resistance in magnetic fields.
(a) Temperature dependence of resistance under $B \bot I$ with different field strengths.
The zero-field curve is also shown.
(b) Arrhenius plot of the data plotted in (a) (omitting the zero-field data).
The solid lines are linear fits.
}
\end{figure}

We now examine the flux-flow resistivity and irreversibility field.
Figure 3(a) plots the resistance as a function of the angle $\theta$ between the magnetic field and current in the resistive-transition region under two conditions: 3.5 T and 0.13 K ($I$ = 7 $\mu$A) and 1 T and 0.69 K ($I$ = 20 $\mu$A).
The observed resistances roughly follow the $\sin^2(\theta)$ curve [broken lines in Figure 3(a)], the expected flux-flow resistance variation in the simple flux-motion model \cite{Kwok90PRL}.
Figure 3(b) compares $R(B)$ curves for the two configurations $B \parallel I$ and $B \bot I$ at $T$ = 0.11 and 0.66 K  ($I$ = 7 $\mu$A).
The clear differences between the two configurations  confirm flux-flow resistance.
We determined the irreversibility fields $B_{irr}$ by applying the criterion of 1\% of $R_N$ to the $B \bot I$ curves and plotted the results in Fig. 2.

Figure 4(a) plots the temperature dependence of the resistance measured at $B \bot I$ ($I$ = 20 $\mu$A) with different field strengths, along with
the zero-field curve for comparison.
The slight difference in the normal-state resistance between $B$ = 0 and $B \geqslant 1$ T is caused by the impurity, which is superconducting at $B$ = 0 but not at $B \geqslant 1$ T.
The transition broadens with increasing applied field.
As shown in Fig. 2, the $B_{irr}$ obtained under the 1\% criterion is consistent with those determined from the $R(B)$ curves in Fig. 3(b). 
Figure 4(b) plots the same data (omitting the zero-field curve) as Arrhenius plots.
Before vanishing, the resistance exhibits thermally activated behavior $R \sim \exp(-E_a/T)$ [solid lines in Fig. 4(b)] resembling the thermally activated flux-flow (TAFF) behavior observed (for example) in cuprates, iron-based superconductors, and superconducting amorphous thin films \cite{Palstra88PRL, Jaroszynski08PRB, Graybeal86PRL}.
The activation energy decreases from 24(2) K at 1 T to 2.7(1) K at 2.5 T.


We now move on to discussion of our results.
First, we compare the superconductivity of the Ta$_{1.6}$Te dd-QC  with that of the Al--Zn--Mg i-QC.
The upper critical field in the Al--Zn--Mg i-QC was determined down to $T/T_c \sim 0.4$ K and was reportedly compatible with WHH theory \cite{Kamiya18NatCommun}. The estimated
$B_{c2}(0)$ was 17 mT, much smaller than the Pauli limit $B_{po} \sim$ 90 mT.
The initial slope of $B_{c2}$, $\mathrm{d}B_{c2}/\mathrm{d}T|_{T_c}$, was approximately $-$0.5 T/K, one order-of-magnitude smaller than that of the Ta$_{1.6}$Te dd-QC ($-$4.4 T/K).
According to weak-coupling BCS theory, the initial slope in the dirty limit is proportional to the product of the normal-state resistivity and the electronic specific-heat coefficient (per volume) \cite{Decroux82Book}.
The resistivity and specific-heat coefficient are one order-of-magnitude and three times larger, respectively, in the Ta$_{1.6}$Te dd-QC \cite{Tokumoto23condmat}  than in the Al--Zn--Mg i-QC \cite{Kamiya18NatCommun}.
This fact crudely explains the much larger $B_{c2}$ slope of the Ta$_{1.6}$Te dd-QC than that of the Al--Zn--Mg i-QC.
Notice that the large slope of $-$4.4 T/K in the Ta$_{1.6}$Te dd-QC can be compared with the initial slopes of $B_{c2}$ in a practical superconductor Nb$_3$Sn ($-2.6$ T/K) \cite{Orlando79PRB}, a heavy-fermion superconductor UPt$_3$ ($-6.3$ T/K) \cite{Chen84PRB}, and an iron-based superconductor (Ba, K)Fe$_2$As$_2$ ($-5.4$ T/K) \cite{Yuan09Nature}, for example.


We now consider the $B_{c2}(T)$ curve.
The linearity of this curve is incompatible with the theoretically proposed FFLO-like state, which gives a $B_{c2}(T)$ curve with an inflection point \cite{Sakai19PRR}.

Two classes of superconductors exhibit approximately linear $B_{c2}(T)$ behavior:
(i) highly disordered systems such as amorphous, metallic glass, and high-entropy alloy superconductors, and (ii) multi-band superconductors such as borocarbide, MgB$_2$, and iron-based superconductors. 

Examples of the former class are the superconducting amorphous alloys (Mo$_{0.5}$Ru$_{0.5}$)$_{80}$P$_{20}$, (Mo$_{0.6}$Ru$_{0.4}$)$_{86}$B$_{14}$, and Mo$_{30}$Re$_{70}$, which give linear $B_{c2}(T)$ versus $T$ curves down to $T/T_c \sim$ 0.2 \cite{Tenhover81SSC}. Si$_{1-x}$Au$_x$ amorphous alloys and Ta--Nb-Hf--Zr--Ti high-entropy alloys \cite{Furubayashi85SSC, Rohr16PNAS} exhibit similar linear $B_{c2}(T)$ curves.
However, it should be noted that linear $B_{c2}(T)$ behavior is comparatively rare in highly disordered superconductors \cite{Karkut83PRB}.
Upward deviations of $B_{c2}(T)$ from the WHH predictions, including $T$-linear behavior, have been arguably ascribed to spatial electronic inhomogeneities on the order of the superconducting coherence length \cite{Carter81SSC}.

The $B_{c2}(T)$ curves of multi-band superconductors exhibit various $T$ dependences ranging from a usual concave one (d$^2B_{c2}$/d$T^2 < 0$) to $T$-linear or convex one \cite{Shulga98PRL, Budko01PRB, Hunte08Nature, Yuan09Nature, Terashima14PRB}.
Theoretically, these variations can be ascribed to different gap sizes between different bands and the relative strengths of intraband and interband scatterings \cite{Gurevich07PhysicaC}.

These examples suggest that departure from the single-uniform-gap picture is a necessary condition of $B_{c2}(T)$ deviation from WHH behavior. More specifically, 
the gap is spatially inhomogeneous in highly disordered superconductors and nonuniform in $k$ space in multi-band superconductors.
Consistent with this conjecture, theoretical studies have shown that the gap in quasicrystal superconductors is intrinsically inhomogeneous \cite{Sakai17PRB, Araujo19PRB, Liu22SCPM}.
However, the sufficient condition of linear $B_{c2}(T)$ is unclear and requires theoretical elucidation.

Next, we discuss why $B_{c2}(0)$ exceeds the Pauli limit by a factor of 2.3.
The Pauli limit is enhanced by electron--phonon coupling as $(1+\lambda_{ep})B_{po}$ \cite{Decroux82Book}.
However, the reported electron--phonon coupling constant in the Ta$_{1.6}$Te dd-QC is $\lambda_{ep}$ = 0.52, which is obviously insufficient to explain the large $B_{c2}(0)$.
We also note that in the presence of both the orbital effect and the Pauli limit, the upper critical field is given by $B_{c2}=B_{c2}^*B_{po}/\sqrt{2(B_{c2}^*)^2+B_{po}^2}$ \cite{Decroux82Book}.
Even when the orbital effect is absent (i.e., $B_{c2}^* \rightarrow \infty$), $B_{c2}$ is limited to $B_{po}/\sqrt{2}$.
Because the experimental $B_{c2}(T)$ curve shows no leveling-off tendency down to 0.04 K, the effective Pauli limit in the Ta$_{1.6}$Te dd-QC must far exceed the experimental $B_{c2}(0)$.

The spin--orbit interaction may be important.
It induces spin-flip scattering and enforces finite spin susceptibility in superconductors even at zero temperature, thus reducing the effect of the Pauli limit.
In the presence of spin-orbit scattering, the effective Pauli limit $B_p$ is enhanced to $B_p = 1.33\sqrt{\lambda_{so}}B_{po}$ \cite{Decroux82Book}.
The spin-orbit scattering parameter $\lambda_{so}$ is given by $\lambda \simeq 1.17\xi_o/l_{so}$, where $\xi_o$ is the BCS coherence length, and $l_{so}$ is the electron mean-free path associated with spin-flip scattering.
Because these parameters are difficult to estimate in the present case, we refer to highly disordered superconductors, in which the $B_{c2}(0)$ sometimes (or likely) exceeds the Pauli limit:
$B_{c2}(0)$ in the superconductors Zr$_{77}$Rh$_{23}$ \cite{Togano75PhysLettA}, Mo$_{45}$Si$_{55}$ \cite{Ikebe81PhysicaB}, Hf$_{80}$Fe$_{20}$ \cite{Tafra08JPCM}, and (TaNb)$_{0.16}$(ZrHfTi)$_{0.84}$ \cite{Rohr16PNAS} exceeds the Pauli limit by $\sim$7, 15, 50, and 9\%, respectively.
These excesses are however much smaller than in the present case.

In noncentrosymmetric superconductors, the spin--orbit interaction causes the antisymmetric spin--orbit interaction and intrinsically suppresses the effect of the Pauli limit.
For instance, when the spin--orbit interaction is the Rashba (Ising) type, the electron spins in noncentrosymmetric crystals are confined to in-plane (out-of-plane) directions.
Therefore, the Pauli limit is removed when the external field is applied perpendicular to the confined direction, and the $B_{c2}$ becomes large \cite{Kimura07PRL, Tada08PRL, Lu15Science, delaBarrera18NatCommun}.

The spin--orbit interaction is stronger in heavier atoms than in lighter atoms, and Ta and Te in our studied quasicrystals are heavy atoms.
Generally, atomic sites in quasicrystals possess no inversion symmetry.
Therefore, we suggest that spin--orbit interactions may play an important role in enhancing the upper critical field in the Ta$_{1.6}$Te dd-QC.

We next consider the influence of possible anisotropy.
The Ta$_{1.6}$Te dd-QC has a layered structure with a vdW gap.
However, the calculated electronic band structure of a crystalline approximant Ta$_{21}$Te$_{13}$ is only moderately anisotropic with sizable dispersion along the out-of-plane direction \cite{Cain20PNAS}.
This finding is likely related to the large thickness of the constituent Te-terminated layers (approximately 1 nm \cite{Conrad02ChemEurJ, Cain20PNAS}),
suggesting that the $B_{c2}$ anisotropy in the Ta$_{1.6}$Te dd-QC, which is determined by the square root of the mass anisotropy, is limited.
Experimentally, the resistive transition under $I \parallel B$ broadens with decreasing temperature [Fig. 1(c)].
This behavior is possibly explained by anisotropy, but the transition width remains a fraction of $B_{c2}$ even at 0.04 K, supporting a small anisotropy.

Finally, we discuss the experimentally observed flux-flow resistance and irreversibility field.
In conventional low-$T_c$ superconductors, the irreversibility field is indistinguishable from the upper critical field.
In contrast, the two critical fields in cuprates and some (relatively) high-$T_c$ or low-dimensional superconductors are distinct and TAFF behavior is observed, highlighting the importance of thermal fluctuations in such superconductors \cite{Blatter94RMP, Gurevich11RPP}.
TAFF-like behavior [Fig. 4(b)], along with distinct irreversibility and upper critical fields (Fig. 2), were observed in the present study.
Thermal fluctuations in the Ta$_{1.6}$Te dd-QC are likely limited by the low $T_c$.
However, the intrinsically inhomogeneous superconducting gap expected in the Ta$_{1.6}$Te dd-QC \cite{Sakai17PRB, Araujo19PRB, Liu22SCPM} can affect the vortex phase diagram and transport properties.
The theoretically suggested peculiar $I-V$ characteristics \cite{Takemori20PRB} and intrinsic pinning sites \cite{Nagai22PRB} must also be considered.

In summary, our resistance and ac susceptibility measurements of the Ta$_{1.6}$Te dd-QC down to 0.04 K revealed the following peculiarities of quasicrystal superconductivity:
The upper critical field increases linearly as the temperature reduces to $T/T_c$ = 0.04, with no leveling-off tendency.
The zero-temperature upper critical field $B_{c2}(0)$ is more than twice the Pauli limit.
The irreversibility field is distinct from the upper critical field.
The flux-flow resistance shows activated behavior.
These observations are possibly explained by the nonuniform gap distribution and spin--orbit interaction in the quasicrystal structure.
To confirm this conjecture, further experimental and theoretical investigations are required.
Lastly, we point out that the large $B_{c2}$ slope ($-$4.4 T/K) is of technological interest.
If large resistivity of quasicrystals tends to cause a large $B_{c2}$ slope, it is worthwhile searching for practical superconductors in quasicrystals.

\begin{acknowledgements}
This study was supported by the JST-CREST program (grant no. JPMJCR22O3; Japan), JSPS KAKENHI (Grant Numbers JP19H05821, JP23K04355, JP22H04485, and JP22K03537), and Tokuyama Science Foundation. MANA is supported by World Premier International Research Center Initiative (WPI), MEXT, Japan.
\end{acknowledgements}

\end{document}


\preprint{ver3}

\title{Supplemental Material\\
Anomalous upper critical field in the quasicrystal superconductor Ta$_{1.6}$Te}



\author{Taichi Terashima}
\affiliation{Research Center for Materials Nanoarchitectonics (MANA), National Institute for Materials Science, Tsukuba 305-0003, Japan}
\author{Yuki Tokumoto}
\author{Kotaro Hamano}
\affiliation{Institute of Industrial Science, The University of Tokyo, Tokyo 153-8505, Japan}
\author{Takako Konoike}
\affiliation{Research Center for Materials Nanoarchitectonics (MANA), National Institute for Materials Science, Tsukuba 305-0003, Japan}
\author{Naoki Kikugawa}
\affiliation{Center for Basic Research on Materials, National Institute for Materials Science, Tsukuba 305-0003, Japan}
\author{Keiichi Edagawa}
\affiliation{Institute of Industrial Science, The University of Tokyo, Tokyo 153-8505, Japan}


\date{\today}

\begin{abstract}
\end{abstract}


\maketitle
\renewcommand{\thefigure}{S\arabic{figure}}


%

\section{Materials and Methods}
The Ta$_{1.6}$Te samples were synthesized via reaction sintering \cite{Tokumoto23condmat}.
A mixture of TaTe$_2$, Ta, and iodine (to promote the reaction) was pressed into a pellet and sintered in a vacuum at 1,273 K for six days.
The synthesis and characterization details are given in \cite{Tokumoto23condmat}.
For the present measurements, one synthesized pellet was chopped in air, and one piece was picked up and roughly shaped with sandpaper.
Electrical contacts were formed with silver conducting paint.

A dilution refrigerator equipped with a 20-T superconducting magnet was used to produce low temperatures down to 0.04 K.
For resistance measurements, the sample was mounted on a top-loading probe and rotated in situ on a rotation platform.
The resistance was measured using the standard four-contact method with a low-frequency ($f \sim$ 17 Hz) ac current.
For ac susceptibility measurements, the electrical wires were removed from the sample and the sample was mounted on another top-loading probe with a pick-up coil. 

\section{Superconducting impurity} 
The present sample contain a small amount of a superconducting impurity whose transition temperature is about 3.2 K [Fig. S1(a)].
The resistance drop associated with the superconducting transition is about 2\% of the sample resistance.
The upper critical field of the superconducting impurity is about 0.47 T at 1.88 K  [Fig. S1(b)].

\begin{figure}
\includegraphics[width=8.6cm]{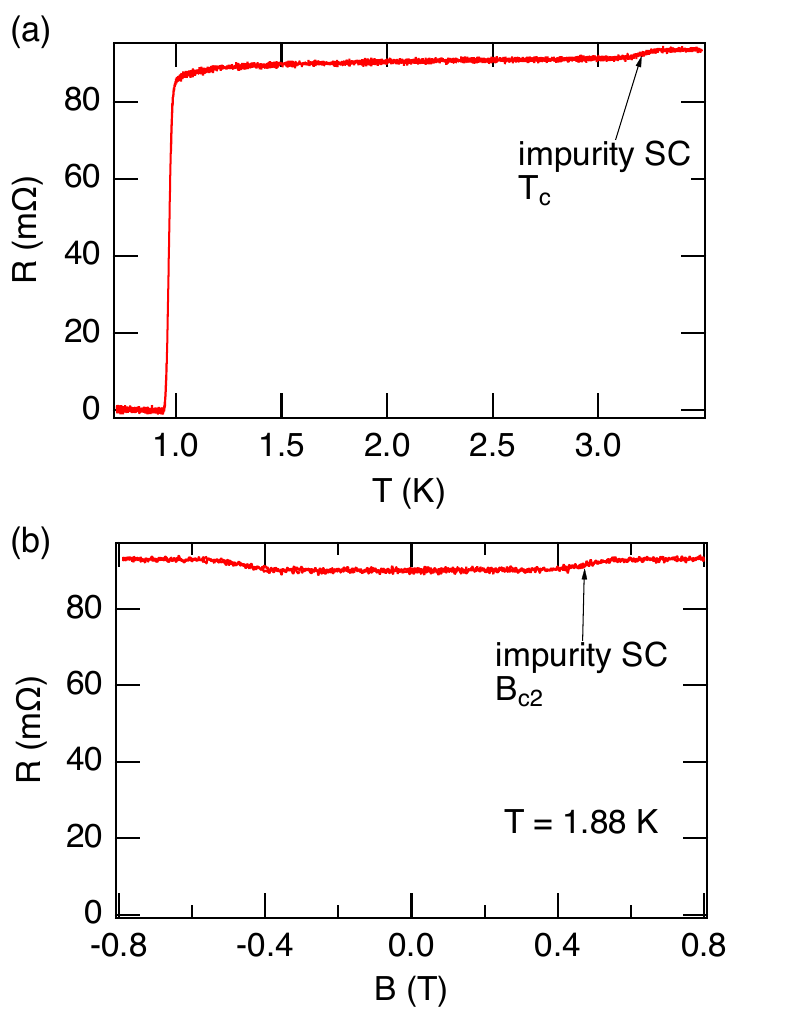}
\caption{\label{Impurity}Superconducting impurity.
(a) Resistance vs temperature curve showing a resistance drop near 3.2 K due to a superconducting impurity.
(b) Resistance vs magnetic field curve measured at 1.88 K showing a resistive transition due to the impurity near 0.47 T.
}
\end{figure}